\documentclass[preprint,12pt,authoryear]{elsarticle}

\usepackage{epsfig}

\usepackage{amssymb}
\usepackage{amsmath}
\usepackage{graphicx}
\usepackage{amsfonts}
\usepackage{xcolor}
\usepackage{caption}
\usepackage{booktabs}
\usepackage{makecell}
\usepackage{comment}
\usepackage{subcaption}
\usepackage{pdflscape}
\usepackage{todonotes}
\usepackage{amsfonts}
\usepackage{booktabs}
\usepackage{multirow}

\journal{Computer Speech \& Language}

\begin{document}

\begin{frontmatter}

%% Title, authors and addresses

%% use the tnoteref command within \title for footnotes;
%% use the tnotetext command for theassociated footnote;
%% use the fnref command within \author or \affiliation for footnotes;
%% use the fntext command for theassociated footnote;
%% use the corref command within \author for corresponding author footnotes;
%% use the cortext command for theassociated footnote;
%% use the ead command for the email address,
%% and the form \ead[url] for the home page:
%% \title{Title\tnoteref{label1}}
%% \tnotetext[label1]{}
%% \author{Name\corref{cor1}\fnref{label2}}
%% \ead{email address}
%% \ead[url]{home page}
%% \fntext[label2]{}
%% \cortext[cor1]{}
%% \affiliation{organization={},
%%            addressline={}, 
%%            city={},
%%            postcode={}, 
%%            state={},
%%            country={}}
%% \fntext[label3]{}

\title{VoxCog: Towards End-to-End Multilingual Cognitive Impairment Classification through Dialectal Knowledge} %% Article title

%% use optional labels to link authors explicitly to addresses:
%% \author[label1,label2]{}
%% \affiliation[label1]{organization={},
%%             addressline={},
%%             city={},
%%             postcode={},
%%             state={},
%%             country={}}
%%
%% \affiliation[label2]{organization={},
%%             addressline={},
%%             city={},
%%             postcode={},
%%             state={},
%%             country={}}

\author[usc_ee]{Tiantian Feng} %% Author name
\author[usc_ee]{Anfeng Xu}
\author[usc_dornsife]{Jinkook Lee}
\author[usc_ee,usc_ling]{Shrikanth Narayanan}

%% Author affiliation
\affiliation[usc_ee]{organization={Ming Hsieh Department of Electrical and Computer Engineering, University of Southern California},%Department and Organization
            addressline={3740 McClintock Ave}, 
            city={Los Angeles},
            postcode={90089}, 
            state={CA},
            country={United States}}
\affiliation[usc_dornsife]{organization={Dornsife College of Letters, Arts and Sciences, University of Southern California},
            addressline={}, 
            city={Los Angeles},
            postcode={90089}, 
            state={CA},
            country={United States}}

%% Abstract
\begin{abstract}
%% Text of abstract
In this work, we present a novel perspective on cognitive impairment classification from speech by integrating speech foundation models that explicitly recognize speech dialects. 
    Our motivation is based on the observation that individuals with Alzheimer's Disease (AD) or mild cognitive impairment (MCI) often produce measurable speech characteristics, such as slower articulation rate and lengthened sounds, in a manner similar to dialectal phonetic variations seen in speech.  
    Building on this idea, we introduce VoxCog, an end-to-end framework that uses pre-trained dialect models to detect AD or MCI without relying on additional modalities such as text or images. Through experiments on multiple multilingual datasets for AD and MCI detection, we demonstrate that model initialization with a dialect classifier on top of speech foundation models consistently improves the predictive performance of AD or MCI. 
    Our trained models yield similar or often better performance compared to previous approaches that ensembled several computational methods using different signal modalities.
    Particularly, our end-to-end speech-based model achieves 87.5\% and 85.9\% accuracy on the ADReSS 2020 challenge and ADReSSo 2021 challenge test sets, outperforming existing solutions that use multimodal ensemble-based computation or LLMs. 
    % Our modeling code is publicly available at: https://github.com/tiantiaf0627/voxcog. (add after accepted)
\end{abstract}

%%Graphical abstract
% \begin{graphicalabstract}
%\includegraphics{grabs}
% \end{graphicalabstract}

%% Keywords

\end{frontmatter}

%% Add \usepackage{lineno} before \begin{document} and uncomment 
%% following line to enable line numbers
%% \linenumbers

%% main text
%%
%% Use \section commands to start a section
\section{Introduction}

Cognitive impairment represents a significant and an increasingly prevalent challenge in the aging population. In particular, Alzheimer's disease (AD) is one of the most common neurodegenerative disorders in the aging population and the leading cause of dementia(~\cite{park2003systematic, ritchie2017midlife, livingston2020dementia}). 
It is characterized by progressive cognitive decline, widespread neuronal degeneration, and eventual neuronal death. 
Apart from AD, Mild Cognitive Impairment (MCI)(~\cite{gauthier2006mild}) indicates a cognitive decline greater than expected for an individual's chronological age. 
Detecting MCI has major implications for the early detection of dementia, as more than 30\% of individuals with MCI develop a type of dementia within 5 years. Since speech is a fundamental neurological function that engages memory, language, and motor planning, there is growing interest in developing speech-based evaluation tasks to identify early signs of cognitive decline and AD. 
For example, DementiaBank(~\cite{lanzi2023dementiabank}) is a widely shared repository that provides speech samples from individuals with different cognitive impairments and healthy controls. It includes data from performing standardized speech tasks, such as the ``Cookie Theft'' picture description or reading task using speech.

While significant efforts have been made in improving Alzheimer's disease predictions, many existing approaches rely on ensembling knowledge from multiple data modalities, e.g., vocal audio, language, to achieve better performance(~\cite{luz21_interspeech, luz2020alzheimer}). 
This adds additional complexity in modeling (e.g., requiring building blocks such as diarization and automatic speech recognition) and results interpretation, and domain experts may find it challenging to understand which modality drives predictions. These additional computational stages also introduce reproducibility challenges by increasing system complexity in implementation details.
Motivated by these challenges, in this work, we pose one research question: \textit{Can an end-to-end speech modeling approach effectively detect cognitive impairment without relying on additional computational stages such as automatic speech recognition (ASR) or language modeling? \textbf{If so, what prior knowledge is informative for this task?}}

To address this question, we investigate whether integrating dialectal speech pattern variations, an information source seemingly orthogonal to cognitive impairment prediction, can improve the detection of MCI or AD. 
Our motivation in using dialectal knowledge is that individuals with cognitive impairment who perform speech tasks may produce speech with varied pronunciations, intonations, and word choices, patterns that are similar to findings in dialect modeling. 
Therefore, based on this observation, we hypothesize that computational models with prior knowledge of dialectal variation can help identify speech markers associated with cognitive impairment. 
Interestingly, our experimental results support that a pretrained dialect classifier provides a strong prior for AD and MCI detection, enabling an end-to-end, speech-only approach that reduces the need to integrate additional modalities such as text or images.

\begin{figure*} {
    \centering
    
    \includegraphics[width=\linewidth]{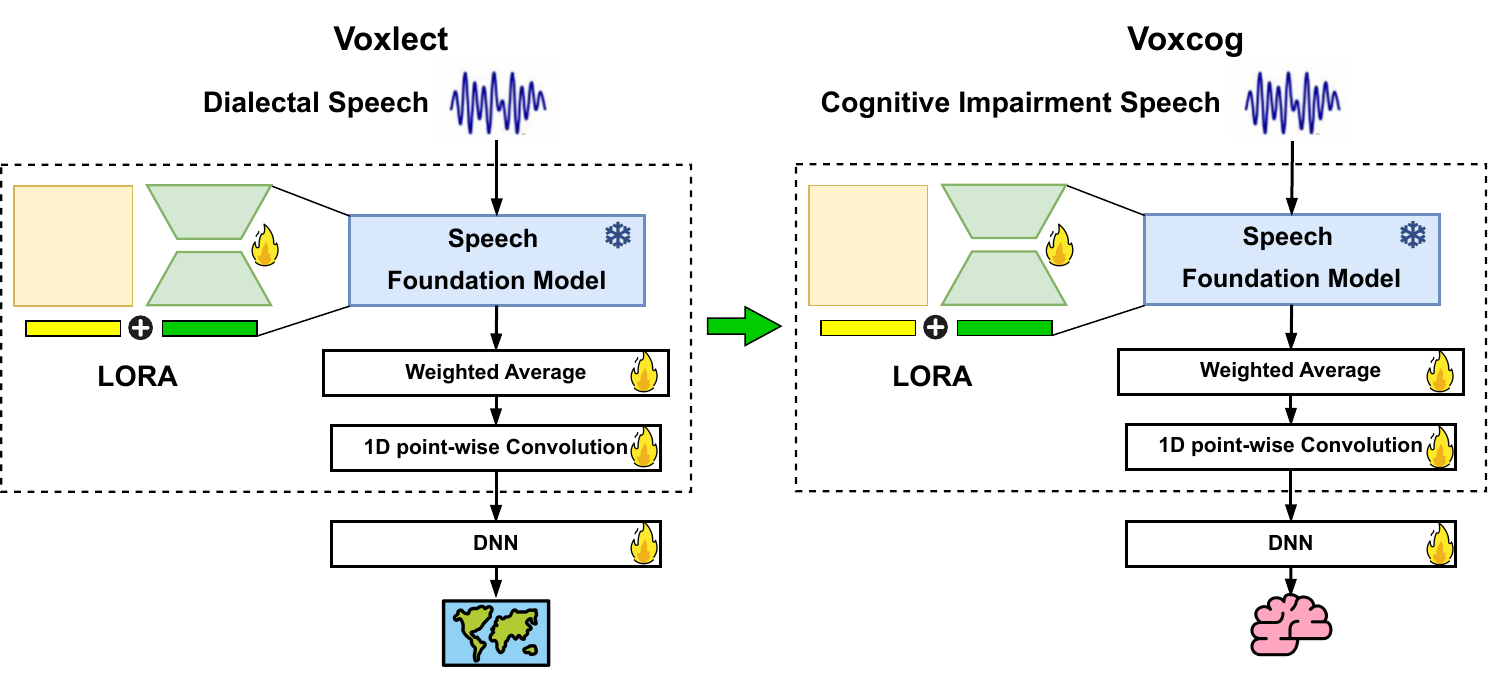}
    
    \caption{Overview of the \texttt{VoxCog} framework. We initialize the VoxCog model training with the model weights from predicting the corresponding dialects. The figure uses sources from the OpenMoji web.}
    \label{fig:voxcog}
} \end{figure*}

Specifically, we make the following contributions in this work: (1) We propose an end-to-end modeling approach, called VoxCog, to classify cognitive impairment that utilizes pre-trained models capturing dialectal knowledge. (2) We report extensive experiments on several widely used cognitive impairment classification datasets involving participants performing various speech tasks, and we show that, in both English and non-English speaking datasets, VoxCog can consistently outperform the baseline backbones that do not explicitly model dialectal information. (3) We further compare VoxCog against representative works from the existing literature and show that it achieves competitive, and in most cases, superior performance.

\begin{figure}[ht] {
    \centering
    
    \includegraphics[width=0.75\linewidth]{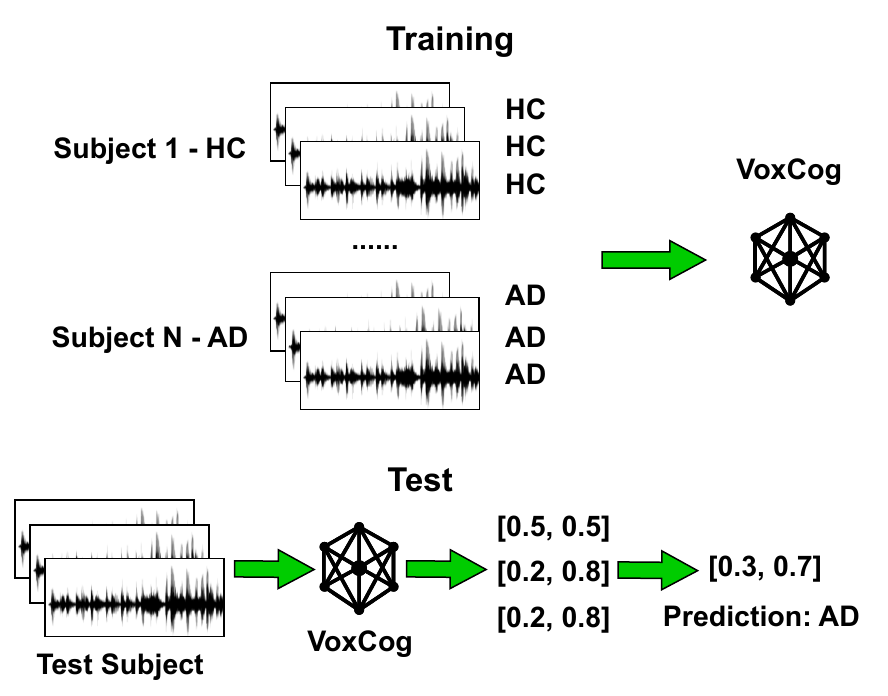}
    \caption{Overview of the data processing in \texttt{VoxCog}.}
    \label{fig:voxcog_data}
} \end{figure}

\section{VoxCog Modeling}
\label{sec:methods}

In this paper, we introduce VoxCog, an end-to-end multilingual framework for cognitive impairment classification that explicitly uses dialectal knowledge learned from large-scale dialectal speech data, as shown in Figure~\ref{fig:voxcog}. The proposed approach initializes the cognitive impairment classification model using weights from a pretrained dialect identification model called Voxlect(~\cite{feng2025voxlect}). Our motivation is to transfer dialectal representations to the downstream task involving cognitive impairment classification. We demonstrate that VoxCog achieves state-of-the-art performance directly using just raw speech signals, without relying on transcriptions or hand-crafted paralinguistic features. These results suggest that dialect-aware speech representations can serve as a data-efficient prior for classifying cognitive health decline across languages.

\subsection{Voxlect}
% \subsection{Dialect Classification}

The dialectal classification models are from our recent Voxlect benchmark, which systematically evaluates the ability of speech foundation models to predict dialectal variation across multiple languages of the world. In this work, we rely on the Voxlect models that predict English, Spanish, and Mandarin dialects. 
We use the pre-trained Whisper-Large(~\cite{radford2023robust}) and MMS-LID-256(~\cite{pratap2024scaling}) as our backbone in training the dialect model.
All models are trained using low-rank adaptation (LoRA)(~\cite{hu2022lora}) and use a unified architecture consisting of a 1D convolutional neural network followed by DNN layers for dialect classification. 
The details of the dialect models are described as follows: 

\subsubsection{English Dialects} Specifically, the English dialect model recognizes 16 global spoken dialects, including those of North America, England, Welsh, Scottish, Northern Irish, Irish, Germanic, Romance, Slavic, Semitic, Oceania, South Africa, Southeast Asia, East Asia, South Asia, and other varieties not mentioned above. The English dialect model was trained using a combination of more than 10 datasets, including TIMIT(~\cite{garofolo1993darpa}), CommonVoice (~\cite{ardila2020common}), EdAcc(~\cite{sanabria2023edinburgh}), British Isles Speaker (~\cite{demirsahin2020open}), L2-Arctic(~\cite{zhao2018l2}), VoxPopuli(~\cite{wang2021voxpopuli}), Fair-Speech(~\cite{veliche2024towards}), ESLTTS(~\cite{wang2024usat}), Hispanic-English(~\cite{hispanic_eng}), ParaSpeechCaps(~\cite{diwan2025scaling}). Utterances shorter than 3 seconds were excluded from dialect modeling, since such segments provide insufficient information for reliable characterization of dialects.

\subsubsection{Spanish Dialects} 
The Spanish dialect model predicts 6 regional dialects: Peninsular, Mexican, Chileno, Andino-Pacífico, Central American and the Caribbean, and Rioplatense. We use the dataset from Latin American Spanish(~\cite{guevara2020crowdsourcing}) and CommonVoice-es(~\cite{ardila2020common}) to train the Spanish dialect model.

\subsubsection{Chinese Dialects}
The Chinese dialect model follows the Mandarin dialect labels provided by KeSpeech(~\cite{tang2021kespeech}). We group samples from Beijing Mandarin and Northeastern Mandarin into Standard Mandarin, given their shared intelligibility with Standard Mandarin. Apart from predicting Mandarin dialects, we include Cantonese samples from CommonVoice(~\cite{ardila2020common}) to enrich the coverage of Chinese languages. Finally, the Chinese dialect model predicts Standard Mandarin, Ji-Lu Mandarin, Southwestern Mandarin, Jiang-Huai Mandarin, Lan-Yin Mandarin, Zhongyuan Mandarin, Jiao-Liao Mandarin, and Cantonese.

\subsection{VoxCog}

\subsubsection{Data Preparation} 

Most speech-based datasets for cognitive impairment classification are collected under constrained experimental settings, in which participants perform a predefined elicitation task (e.g., picture description). This results in recordings of relatively short duration, typically on the order of 1–2 minutes. To accommodate this recording condition, we segment each subject's recording using a sliding window of 15 seconds with a 5-second step size and assign the corresponding cognitive-impairment label to each segment for training. This is motivated by the relatively short duration of the recordings (approximately 1–2 minutes), for which 15-second speech segments include sufficient information for classifying cognitive impairment. At inference time, segment-level predictions are aggregated to produce an individual subject-level cognitive impairment prediction. The overview of data processing and inference time aggregation is presented in Figure~\ref{fig:voxcog_data}.

\subsubsection{Model Architecture} 

We initialize the model using Voxlect pretrained weights(~\cite{feng2025voxlect}) adapted with LoRA modules and 1D point-wise convolutional layers, while the downstream DNN layers are randomly initialized for cognitive impairment classification. During fine-tuning, we unfreeze the pretrained LoRA and convolutional layers jointly with the DNN classifier, which enables task-specific adaptation for cognitive impairment classification.

\section{Datasets}
\label{sec:dataset}

We experimented with 6 speech-based datasets to predict cognitive impairment, including AD or MCI status, as described in Table~\ref{tab:datasets}. All datasets, except for 2021-NCMMSC(~\cite{chen2023raw}), were downloaded from the DementiaBank(~\cite{lanzi2023dementiabank}) website.

\subsection{2020 ADReSS Challenge Dataset}

Our first experimental dataset is the INTERSPEECH 2020 ADReSS Challenge dataset(~\cite{luz2020alzheimer}). This dataset was designed with a balanced distribution of AD and healthy control (HC) subjects. The training set contains 108 subjects, with 54 individuals with AD and 54 healthy controls. Similarly, the test set includes 48 subjects, evenly split between the two groups. Speech data were collected from English-speaking subjects performing a picture-description task based on the Cookie Theft image(~\cite{kaplan2001boston}).

\begin{table}
    \centering

    \caption{Summary of group statistics in the datasets used in this work. (HC: Healthy Control; MCI: Mild Cognitive Impairment; AD: Alzheimer's Disease; DM: Dementia)}
    % DM: Dementia
    \vspace{0.5mm}
    \resizebox{\linewidth}{!}{
    \begin{tabular}{lccc}

        \toprule
        \textbf{Dataset} & 
        \multicolumn{1}{c}{\textbf{Language}} & 
        \multicolumn{1}{c}{\textbf{Groups}} & 
        \multicolumn{1}{c}{\textbf{Task}} 
        \\
        \midrule

        ADReSS &
        English & 
        78 HC, 78 AD & 
        Description Task
        \\

        ADReSSo &
        English & 
        115 HC, 122 AD & 
        Description Task
        \\

        VAS &
        English & 
        30 HC, 30 DM & 
        Interaction with Alexa
        \\

        % TAUKADIAL-en & English & 63 HC, 123 MCI &  Description \\

        \midrule

        2021-NCMMSC &
        Chinese & 
        44 HC, 26 MCI & 
        Description Task
        \\

        TAUKADIAL-zh &
        Chinese & 
        43 HC, 44 MCI & 
        Description Task
        \\

        Ivanova &
        Spanish & 
        197 HC,91 MCI,74 AD & 
        Reading
        \\

        \bottomrule

    \end{tabular}
    }
    \label{tab:datasets}
\end{table}

\subsection{2021 ADReSSo Challenge Dataset}

The second dataset used in this work is the ADReSSo Challenge dataset, which was originally released as part of the INTERSPEECH ADReSSo 2021 Challenge(~\cite{luz21_interspeech}). The complete dataset includes speech recordings from 237 participants, of which 166 subjects are in the training set, and 71 subjects are in the test set. Among these participants, 87 were labeled as AD subjects, while 79 were healthy control subjects. Similar to the 2020 ADReSS Challenge Dataset, all participants were native English speakers and were instructed to complete a picture-description task using the Cookie Theft picture(~\cite{kaplan2001boston}).

\subsection{VAS Dataset}

Our third experimental dataset uses the VAS corpus(~\cite{kurtz2023early}) that was collected by asking participants to interact with the Alexa Echo device. 
The data used in this analysis were collected remotely due to the COVID-19 pandemic. In the experiments of this paper, we focus on classifying HC and AD subjects.

\subsection{Spanish Ivanova Dataset}

The first non-English dataset we experimented with in this paper is the Spanish Ivanova Dataset(~\cite{ivanova2022discriminating}). The dataset consists of speech recordings from 361 native Spanish speakers. There is a total of 74 individuals with AD, 90 individuals with MCI, and the remaining 197 individuals are healthy control subjects. All subjects were instructed to read the first two sentences of Don Quixote by Cervantes.

\subsection{2021-NCMMSC Dataset}

The second non-English experimental dataset uses the 2021-NCMMSC Challenge Dataset(~\cite{chen2023raw}). In particular, we focus on detecting AD from HC subjects in this experiment, which includes 26 AD and 44 HC subjects. Most participants are native Mandarin speakers and were instructed to perform picture description, fluency test and free conversation in the recording.

\subsection{TAUKADIAL-zh Dataset}

The last dataset we used is the TAUKADIAL challenge dataset, which focuses on classifying Mild Cognitive Impairment (MCI) and healthy participants. The dataset includes speech recordings %of participants 
involving tasks such as picture descriptions. The dataset contains native speakers in either English or Chinese. We use the Whisper-large to infer the language used in each recording and select the Chinese-speaking subset of the dataset to evaluate VoxCog as part of the non-English speaking data evaluation.

\section{Experiments}
\label{sec:experiments}

\subsection{Data Split} Specifically, we perform 3-fold cross-validation on the 2020 ADReSS, TAUKADIAL-zh, and 2021 NCMMSC Challenge datasets, while we perform 5-fold cross-validation on the 2021 ADReSSo Challenge dataset.
For ADReSS, ADReSSo, and NCMMSC datasets, we report the average macro-F1 score and accuracy on the held-out test sets. We compute the unweighted average recall (UAR) for the TAUKADIAL challenge dataset. These metrics were selected in accordance with the respective challenge rules. The fold numbers were heuristically chosen based on the number of subjects in each dataset. 
Furthermore, we conduct 5-fold cross-validation on the Spanish Ivanova dataset and the VAS dataset, and report the average of Macro-F1 and accuracy across test folds. For all experiments, we report subject-level prediction scores.

\subsection{Training Details} In our training, we apply a set of data augmentations, including adding Gaussian noise, background noise, time stretching, and polarity inversion, to improve model robustness. We fine-tune both the original and Voxlect-adapted variants of the Whisper-Large(~\cite{radford2023robust}) and MMS-LID-256(~\cite{pratap2024scaling}) models in our experiments. All experiments are trained for up to 10 epochs, with the learning rate selected from [0.0001, 0.0002, 0.0005, 0.001, 0.002]. 
% We chose to train 5 epochs for the TAUKADIAL Dataset due to its faster converging speech. 

\begin{table*}
    \centering

    \caption{Comparison of cognitive impairment classification between VoxCog and the baseline method in the English dataset. The average macro-F1 and accuracy across different training and validation folds are reported.}
    % \vspace{1mm}
    \resizebox{\linewidth}{!}{
    \begin{tabular}{lccccccc}

        \toprule

        \multirow{2}{*}{\textbf{Dataset}} & 
        \multirow{2}{*}{\textbf{Prediction}} & 
        \multicolumn{2}{c}{\textbf{Whisper-Large}} & \multicolumn{2}{c}{\textbf{VoxCog-Whisper-Large}} 
        \\
         & & 
        \multicolumn{1}{c}{\textbf{Macro-F1}} & \multicolumn{1}{c}{\textbf{Acc}} & \multicolumn{1}{c}{\textbf{Macro-F1}} & \multicolumn{1}{c}{\textbf{Acc}} 
        \\
        \cmidrule(lr){1-2}
        \cmidrule(lr){3-4}
        \cmidrule(lr){5-6}

        2020-ADReSS &
        HC $/$ AD & 
        0.801 & 
        80.55 & 
        \textbf{0.846} & 
        \textbf{84.72} 
        \\

        2021-ADReSSo &
        HC $/$ AD & 
        0.763 & 
        76.61 & 
        \textbf{0.819} & 
        \textbf{81.69} 
        \\

        VAS &
        HC $/$ DM & 
        0.852 & 
        86.21 & 
        \textbf{0.913} & 
        \textbf{91.51} 
        \\

        \midrule

        \multirow{2}{*}{\textbf{Dataset}} & 
        \multirow{2}{*}{\textbf{Prediction}} & 
        \multicolumn{2}{c}{\textbf{MMS-LID-256}} & \multicolumn{2}{c}{\textbf{VoxCog-MMS-LID-256}}
        \\

        & & \multicolumn{1}{c}{\textbf{Macro-F1}} & \multicolumn{1}{c}{\textbf{Acc}} & \multicolumn{1}{c}{\textbf{Macro-F1}} & \multicolumn{1}{c}{\textbf{Acc}} 
        \\
        \cmidrule(lr){1-2}
        \cmidrule(lr){3-4}
        \cmidrule(lr){5-6}

        2020-ADReSS &
        HC $/$ AD & 
        0.704 & 
        70.83 & 
        \textbf{0.705} & 
        \textbf{70.83}
        \\
        
        2021-ADReSSo &
        HC $/$ AD & 
        0.638 & 
        64.22 & 
        \textbf{0.696} & 
        \textbf{69.85}
        \\

        VAS &
        HC $/$ DM & 
        0.896 & 
        89.84 & 
        \textbf{0.896} & 
        \textbf{90.00}
        \\

        \bottomrule

    \end{tabular}
    }
    \label{tab:voxcog_en_results}
\end{table*}

\section{Results}

\subsection{Can VoxCog outperform baselines on English data?}

We first compare our proposed VoxCog models against baseline models that do not explicitly model the English-speaking dialects, as summarized in Table~\ref{tab:voxcog_en_results} for the English-language datasets. Overall, we observe that VoxCog can consistently outperform corresponding baseline models regardless of the backbone models used. For example, the VoxCog–Whisper-Large model achieves average accuracies of 84.72\% and 81.69\% on the 2020 ADReSS Challenge and 2021 ADReSSo Challenge datasets, respectively. In contrast, the baseline Whisper-Large model only yields accuracies of 80.55\% and 76.61\% on the same datasets. These results suggest that explicitly modeling spoken English dialectal characteristics can lead to improved accuracy in cognitive impairment detection. 

\subsection{Performance of VoxCog on non-English data}

In addition to evaluating VoxCog on English-speaking datasets, we compare VoxCog with baseline models on non-English speech datasets, including Spanish and Chinese. In particular, results in Table~\ref{tab:voxcog_other_results} show that VoxCog can consistently outperform the baseline models across these multilingual settings. For example, on the Ivanova dataset, VoxCog achieves a Macro-F1 score of 0.738 for HC versus AD classification, improving the baseline performance of 0.694. Similarly, on the Chinese subset of the TAUKADIAL dataset, VoxCog improves the Macro-F1 score from 0.571 to 0.603 compared to the baseline.
These results suggest that the proposed VoxCog can generalize effectively across language settings.

\begin{table}
    \centering

    \caption{Comparing VoxCog with baseline model in classifying cognitive impairment on non-English datasets. The unweighted average recall is reported for TAUKADIAL, and the Macro-F1 score is used for Ivanova and the 2021-NCMMSC dataset.}
    % \vspace{1mm}
    \resizebox{0.9\linewidth}{!}{
    \begin{tabular}{lcccc}

        \toprule

        \multirow{2}{*}{\textbf{Dataset}} & 
        \multirow{2}{*}{\textbf{Prediction Group}} & 
        \multirow{2}{*}{\textbf{Whisper-Large}} & \multicolumn{1}{c}{\textbf{VoxCog}}
        \\

        & & & \multicolumn{1}{c}{\textbf{Whisper-Large}}
        \\

        \cmidrule(lr){1-2}
        \cmidrule(lr){3-4}

        2021-NCMMSC &
        HC $/$ AD & 0.854
         & \textbf{0.940}
        \\

        TAUKADIAL-zh &
        HC $/$ MCI & 0.571
         & \textbf{0.603}
        \\
        
        Ivanova &
        HC $/$ MCI & 
        \textbf{0.631} & 
        {0.626} 
        \\

        Ivanova &
        HC $/$ AD & 
        {0.715} & 
        \textbf{0.718} 
        \\
        \midrule
        
        \multirow{2}{*}{\textbf{Dataset}} & 
        \multirow{2}{*}{\textbf{Prediction}} & 
        \multirow{2}{*}{\textbf{MMS-LID-256}} & \multicolumn{1}{c}{\textbf{VoxCog}}
        \\

        & & & \multicolumn{1}{c}{\textbf{MMS-LID-256}}
        \\
        
        \cmidrule(lr){1-2}
        \cmidrule(lr){3-4}

        2021-NCMMSC &
        HC $/$ AD & 0.894
         & \textbf{0.952}
        \\
        
        TAUKADIAL-zh &
        HC $/$ MCI & 0.517
         & \textbf{0.520} 
        \\
        
        Ivanova &
        HC $/$ MCI & 
        0.585 & 
        \textbf{0.634} 
        \\

        Ivanova &
        HC $/$ AD & 
        0.694 & 
        \textbf{0.738} 
        \\
        
        \bottomrule

    \end{tabular}
    }
    \label{tab:voxcog_other_results}
\end{table}

\subsection{Comparing VoxCog with Existing Literature}

We further benchmark the proposed VoxCog model against representative approaches reported in the literature. To ensure a fair comparison, we restrict our evaluation to the ADReSS 2020 and ADReSSo 2021 challenge datasets, both of which provide standardized test sets that ensure cross-study comparison. Given the challenge-based nature of these datasets, many presented works often apply ensemble approaches to optimize performance. Similarly, we adopt a simple ensemble where we aggregate predictive probabilities from models trained on different cross-validation folds. This aligns our evaluation with common practices in the challenge literature while avoiding overly complex ensemble designs.

\begin{table}
    \centering

    \caption{Comparing VoxCog with representative systems reported in the existing literature. The reported numbers are based on average model predictions trained on different folds. Particularly, boldface denotes the best-performing result for each metric, while underlining indicates the second-best result.}
    % \vspace{1mm}
    \resizebox{\linewidth}{!}{
    \begin{tabular}{lcccccccccccc}

        \toprule

        \multicolumn{1}{c}{\textbf{Dataset}} & \multicolumn{1}{c}{\textbf{Model}} & 
        \multicolumn{1}{c}{\textbf{Macro-F1}} & \multicolumn{1}{c}{\textbf{Accuracy}} 
        \\
        \cmidrule(lr){1-1}
        \cmidrule(lr){2-2}
        \cmidrule(lr){3-4}

        \multirow{5}{*}{2020-ADReSS} & Challenge Baseline(~\cite{luz2020alzheimer}) & 0.745 & 75.00
        \\
        
        & INESC-ID(~\cite{pompili2020inesc}) & 0.808 & 81.25
        \\

        & JHU x-vector(~\cite{pappagari2020using}) & 0.745 & 75.00
        \\

        & RMIT System(~\cite{syed20_interspeech}) & - & \underline{85.42}
        \\

        & \textbf{VoxCog (Ours)} & {\textbf{0.875}} & {\textbf{87.50}}
        \\ \midrule

        \multirow{6}{*}{2021-ADReSSo} & Challenge Baseline(~\cite{luz21_interspeech}) & 0.789 & 78.87
        \\
        
        & PPGs-BERT(~\cite{sun2025ppgs}) & - & 73.20
        \\

        & Whisper-MJT(~\cite{jia2025whisper}) & - & 74.65
        \\

        & MUET-RMIT(~\cite{syed2021tackling}) & \underline{0.845} & \underline{84.51}
        \\

        & WavBERT(~\cite{zhu2021wavbert}) & 0.830 & 83.10
        \\

        & CogBench(~\cite{feng2025cogbench}) & 0.661 & 66.48
        \\
        
        & \textbf{VoxCog (Ours)} & \textbf{0.859} & \textbf{85.92}
        \\ 
        \bottomrule

    \end{tabular}
    }
    \label{tab:voxcog_sota}
\end{table}

\vspace{1mm}
\noindent \textbf{2020-ADReSS}
The results in Table~\ref{tab:voxcog_sota} show that our proposed VoxCog achieves a competitive accuracy of 87.5\% with representative approaches reported in the literature, including challenge submissions that developed ensembles of acoustic and language-based models. In particular, VoxCog outperforms the majority of baselines that rely predominantly on acoustic representations, such as x-vector-based systems(~\cite{pappagari2020using}). It is important to note that VoxCog is a unified end-to-end framework that requires only raw speech signals as input, which simplifies both the training and inference pipelines.

\vspace{1mm}
\noindent \textbf{2021-ADReSSo}
On the larger 2021-ADReSSo dataset, the proposed VoxCog model also consistently outperforms all reported systems, even where many of those involve ensembling language representations in their modeling. Specifically, VoxCog outperforms WavBERT(~\cite{zhu2021wavbert}), which ensembles both semantic and non-semantic linguistic information. Notably, VoxCog can also outperform the LLM baselines reported in VoxBench(~\cite{feng2025cogbench}).
This suggests that the VoxCog model can effectively achieve competitive performance with top-performing solutions in the literature without additional computations, like ASR or language modeling.

\section{Conclusion}

In this work, we introduce {VoxCog}, an end-to-end modeling approach that incorporates dialectal knowledge to improve the prediction of cognitive impairment from speech. In particular, {VoxCog} builds upon our prior large-scale benchmarking efforts in dialect classification across multiple languages. We evaluate the proposed approach on several widely used speech-based cognitive impairment classification datasets across different languages. Experimental results show that integrating pre-trained dialectal knowledge consistently improves performance over baseline models that do not explicitly model speaking dialects. 
This empirically validates our hypothesis that dialectal data carry systematic phonetic, prosodic, and articulatory patterns of a language, which are informative for detecting atypical speaking behaviors in individuals with cognitive decline, whose speech often deviates from normative speaking patterns.
Furthermore, the proposed end-to-end {VoxCog} framework achieves performance that is competitive with representative works reported in the literature. One practical challenge in the current recordings is the presence of the interviewer's speech. In future work, we plan to integrate robust speaker diarization techniques to more accurately extract the speech regions of interest before downstream classification to reduce potential confounds.

\section{Acknowledgment}
\label{sec:ack}

This work is funded by NIA/NIH (R01 AG080473) and NIH (R01 AG051125).

%% If you have bib database file and want bibtex to generate the
%% bibitems, please use
%%
\bibliographystyle{elsarticle-harv} 
\bibliography{main}

%% else use the following coding to input the bibitems directly in the
%% TeX file.

%% Refer following link for more details about bibliography and citations.
%% https://en.wikibooks.org/wiki/LaTeX/Bibliography_Management

\end{document}